\documentclass[conference]{IEEEtran}
\usepackage{eso-pic}
\usepackage{graphicx}
\usepackage{amssymb}

\begin{document}

\title{Variation principle and the universal metric of dynamic routing}

\author{\IEEEauthorblockN{A.M. Sukhov}
\IEEEauthorblockA{Samara State Aerospace University\\Moskovskoe sh. 34,\\
Samara, 443086, Russia\\
{\it e-mail: amskh@yandex.ru}}
\and
\IEEEauthorblockN{D.Yu. Chemodanov}
\IEEEauthorblockA{Samara State Aerospace University\\Moskovskoe sh. 34,\\
Samara, 443086, Russia\\
{\it e-mail: duman190@gmail.com}}}

\maketitle

\begin{abstract}
In this paper the variation principles from theoretical physics is considered that would describe the process of routing in computer networks. The total traffic which is currently served on all hops of the route has been chosen as the quantity to minimize. Universal metric function has been found for dynamic routing taking into account the packet loss effect. An attempt to derive the metric of the most popular dynamic routing protocols such as RIP, OSPF, EIGRP from universal metric was made.
\end{abstract}

\section{Introduction}
\label{intr}

In today's networks different routing principles are used, the most time-consuming method of determining the route is a static method. Taking into account the fact that the number of routing nodes in the Internet is growing and exceeds tens of billions of routers, administrators cannot confine to static routing. Search route in today's networks is a dynamic manner based on the metric. In order to calculate it different algorithms (Dijkstra's algorithm~\cite{knuth,8}, a distance-vector algorithm~\cite{9,wpd}, the optimized link state routing~\cite{10,11}) are used forming the basis for the basic dynamic routing protocols such as RIP, OSPF, EIGRP~\cite{1}. 

Despite the large number of existing algorithms, the question of constructing a universal metric has not yet been resolved in general form. An analytical solution for the description of packet forwarding (routing) is not found. Currently, the process of route selection is to find appropriate metric characterizing features of the route. Subsequently, its values for different routes are compared, and the route with the minimum value of the metric is chosen. 

The aim of this work is to find a universal metric for dynamic routing based on the variation principle. Different areas of physics have long used the concept of extremum for specially selected quantity:
\begin{itemize}
	\item 
In classical mechanics it is called the principle of least action.
	\item 
Similarly, extremal principles have formulated in classical electrodynamics and general theory of relativity, with the only difference that now to the action of particles in external fields is added to the action, describing the variation of fields themselves. In quantum electrodynamics and quantum gravity, these principles also follow from the corresponding functional integrals.
	\item 
In geometrical optics Fermat's principle plays this role (the principle of least time).
	\item 
In thermodynamics the quantity for which the extremum is sought, is the entropy.
\end{itemize}
In theoretical physics there is a course~\cite{2} where all the basic equations are derived for all areas based on the extremal principle.

This article is organized as follows: Section~\ref{sb} substantiates the choice of the universal metric functions, in Section~\ref{s3} attempts is made to derive the metrics of basic dynamic protocols based on a universal metric.

\section{Action selection}
\label{sb}

From all the physical principles the closest one to telecommunications is the principle of Fermat.

Fermat's principle in geometrical optics is a postulate, directing the ray of the light to move from the starting point to end point on the path which minimizes the optical path length:
\begin{equation}
	\delta\int\limits_A^B n dl=0,
	\label{1}
\end{equation}
or for areas with constant refractive index:
\begin{equation}
	S=\min \sum\limits_i n_i l_i,
		\label{2}
\end{equation}
where $l_i$ is the distance traveled by light in a medium with a refractive index $n_i$.

This principle is chosen as the starting point for describing routing in computer networks.

The second starting point to determine the universal function is the law of Little's queuing theory. The classical formulation of this theorem states that the average number of customers $N$ in the stably functioning system is the product of the average arrival rate of customer  $\lambda$ by the average time $T$ they spend in the system:
\begin{equation}
		N=\lambda T
		\label{3}
\end{equation}

\begin{figure}
\centering
\includegraphics[width=9cm]{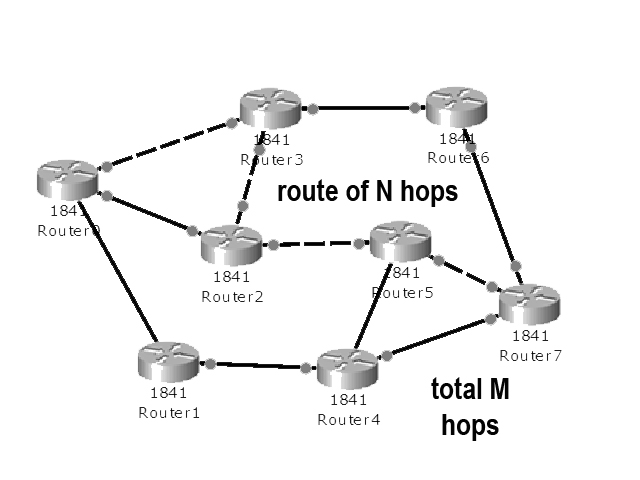}
\caption{Routing scheme}
\label{fi1}
\end{figure}

In this paper the notations the basic characteristics of the network adopted in the works are used~\cite{3,4,5}. Suppose we are given a network consisting of $M$ routers connected communication channels (see Fig.~\ref{fi1}). Then an arbitrary route consists of $N$ hops. Here and below, the following definitions:
\begin{itemize}
	\item 
The hop is a connection of the  "point to point", it includes a router and a data channel that connects the router to another one.
		\item 
The route is a set of hopes at which packets are sent from the sender to the recipient.
\end{itemize}

\begin{figure}
\centering
\includegraphics[width=7cm]{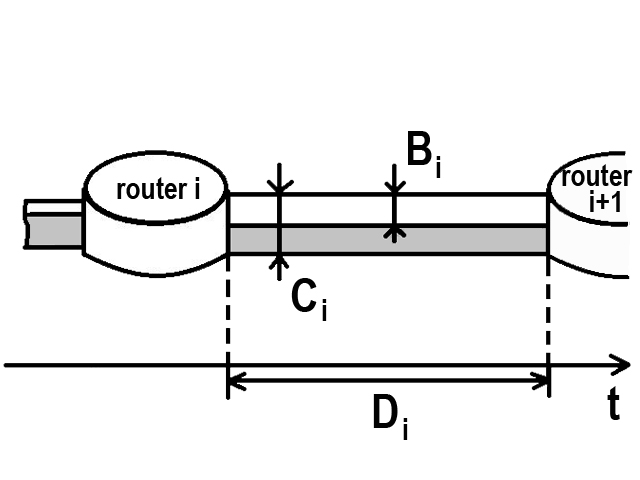}
\caption{Illustration of the basic quantities}
\label{fi2}
\end{figure}

Arbitrary hop is characterized by the following values (see Fig.~\ref{fi2}):

\begin{itemize}
	\item 
The capacity $C_i$ is the maximum IP-layer throughput that the path can provide to a flow, when there is no competing traffic load. 
		\item 
The available bandwidth $B_i$ is the maximum IP-layer throughput that the path can provide to a flow, given the path's current cross traffic load.
		\item 
$D_i$, delay is the time at which the packet is transmitted over the hop
		\item 
$p_i$, percentage of losses is the amount of lost packets to the total number of transmitted packets.
		\item 
$D_i^{buf}$, duration of dejitter buffer period of time within which it is necessary to save the packets on the receiving side of the router in order to eliminate the effect of irregularity delivery of the package. This is the time between the maximum allowable waiting time in queues and averaged value of delays. Network jitter $j$ is defined as the variation of the delay.
\end{itemize}

The Little's law allows finding traffic, which is currently served by the system on the testing route in Fig.~\ref{fi1}:
\begin{equation}
	W=\sum\limits_{i=1}^N (C_i-B_i)D_i,
	\label{f4}
\end{equation}
where the summation is over all parts of the route.

However, this function does not take into account the impact of packet loss $p_i$ on the network hop. In order to describe this effect we note that the packet is considered lost if the time of delivery exceeds a certain time interval $D^{max}$. Therefore, the metric function of the Eqn.~(\ref{f4}) should introduce an additional term:
\begin{equation}
	W=\sum\limits_{i=1}^N (C_i-B_i)((1-p_i)D_i + p_i D^{max}),
	\label{f5}
\end{equation}
or this equation can be reduced to:
\begin{equation}
	W=\sum\limits_{i=1}^N (C_i-B_i)(D_i + p_i D_i^{buf}),
	\label{f6}
\end{equation}
where $D_i^{buf}=D^{max}-D_i$ is the time to allow smooth the effect delay variation (network jitter)~\cite{6}. Let try to estimate the dependence of the dejitter buffer $D_i^{buf}$ on the value of the jitter $j$. For this assessment the percentage of packet loss, which should be due to delay variation, should be estimated. The corresponding expression can be derived from a Poisson distribution, or knowing the type of delay distribution~\cite{7}. Both of these approaches will lead to the expression
\begin{equation}
	p=e^{-\lambda D^{buf}},
	\label{f7}
\end{equation}
where $\lambda=1/j$. We obtain for the dejitter buffer 
\begin{equation}
	D^{buf}=-j \ln p,
	\label{f8}
\end{equation}
for more specific routing problem it can be put $D^{buf}=-7j$. This estimation is based on the assumption that the overall percentage of packet loss is less than 0.5\%~\cite{cal} and only 20\% of data loss due to dejitter effect~\cite{sag}.

Function from Eqn.~\ref{f6} can be chosen as an analogue of action from the physics. Minimum condition for this quantity allows us to select the best route:
\begin{equation}
	\min\limits_{j=\overline{1,K}} W_j=\min\limits_{j=\overline{1,K}} \sum\limits_{i=1}^N (C_i-B_i)(D_i + p_i D_i^{buf}),
	\label{f9}
\end{equation}
where $K$ is the number of comparable routes.

In the next section an attempt will be made to derive the basic metrics of routing protocols from this universal metric.

\section{Metrics of major routing protocols}
\label{s3}

Engineers have not had time for detail study during the explosive growth of global networks, it was necessary to quickly create working routing algorithms. Therefore, initially the simplest algorithms were created on the basis of simple experiments, which subsequently became more sophisticated. Thus there is one additional condition; new algorithms should be based on old metrics in limiting cases. So these metric functions often contain only the linear terms of the complex dependencies of routing, which have been described in Eqn.~(\ref{f9}). In this section, the main function used in routing protocols, will be obtained by simplifying the universal metric.

\begin{figure}
\centering
\includegraphics[width=7cm]{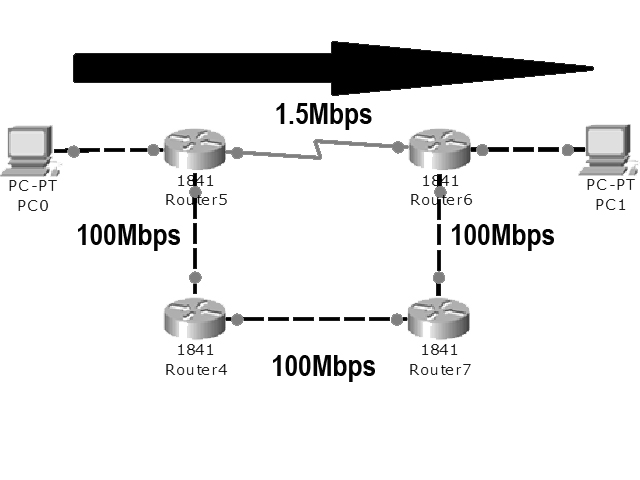}
\caption{Illustration of routing for RIP metrics}
\label{fi3}
\end{figure}

The simplest dynamic routing protocol is {\it RIP} (Routing Information Protocol). Number of hops is used as the metrics. Below, in Fig.~\ref{fi3} it is illustrated how the {\it RIP} protocol comes in choosing a route: instead of three more sections of high-speed backbone {\it RIP} selects the best route for a hops number, though much worst for speed.

Put in a universal routing metric from Eqn.~(\ref{f6}): $B_i=B$, $C_i=C$, $D_i=D$, $p_i=0$, then:
\begin{equation}
	W=(C-B)DN,
	\label{f10}
\end{equation}
where $N$ is number of hops, minimization procedure leads to the choice of the route with minimum number of hops.

{\it OSPF} (Open Shortest Path First) is dynamic routing protocol based on link-state technology and are using Dijkstra's algorithm for finding the shortest path. 

{\it OSPF} allows for optimum bandwidth utilization. Metric determines the weight of edge of the graph, as $10^6/B_i$.  Total metric function $W_j=\sum_{i=1}^N 10^6/B_i$  is constructed for comparison the possible $K$ routes, and then the route with the lowest value $W_j$ of the metric is chosen:
\begin{equation}
	W=\min\limits_{j=\overline{1,K}} \sum\limits_{i=1}^N \frac{10^6}{B_i}
	\label{f11}
\end{equation}

The universal metric from Eqn.~\ref{f9}, provided $D_i=D$, $p_i=0$, is described by a decreasing affine function. As mentioned above this dependency (from $B_i$) they tried to establish an experimental way, it was erroneously described the dependence  $1/B_i$, and that is reflected in the metric of  {\it OSPF}. It should be noted that the only simplest dependences are investigated by the interpretation of experimental results, and replacement of an affine function on the inverse function does not lead to a strong distortion of the results for a small number of obtained points.

{\it EIGRP} metric consists of five basic variables (the default is only two):
\begin{itemize}
	\item 
$bandwidth=\frac{10^6}{B} 256$, where $B=\min B_i$ is the smallest available bandwidth between two points investigated (the default);
	\item 
$delay=D*256$, $D$ is the total delay on the route ($D=\sum_{i=1}^N D_i$); 
		\item 
$relability=256*p$ is worst reliability score all the route, depends on the packet loss in our notation $p=\max p_i$; 
		\item 
$loading=256*l$ is channel utilization rate, where $l=(C_i-B_i)/C_i$; 
		\item 
$MTU$ is smallest {\it MTU} (Maximum Transmission Unit) on the path.
\end{itemize}

By default, the variable bandwidth and delay are used to calculate metric with some weight coefficients. The remaining metric functions are not recommended, as this will result in frequent route recalculation. EIGRP counts a metric using two types of coefficients, the weight and base ones. Basic values of the coefficients by default are: $K1=K3=1$, $K2=K4=K5=0$. 

If the values of the coefficients are equal to the basic default values, taking into account the weighting coefficients metric function will be as
\begin{equation}
	W=bandwidth + delay
	\label{f12}
\end{equation}

It should be noted that under the construction of this metric it was necessary to take into account the experience of the {\it OSPF} protocol, and to reflect the universal dependence of the metric function of packet delay $D$, as well as other quantities appearing in Eqn.~(\ref{f9}). As a result,  {\it EIGRP} appeared. In order to reproduce the dependence found in a universal metric, imposition of weights needed.

The delay in Eqn.~(\ref{f12}) is indicated in tens of microseconds ($10^{-6}$ {\it s}, microseconds). The delay showed by the commands {\em ip eigrp topology} or {\em show interface}, specified in microseconds, respectively, this value should be divided by 10 before using in this formula. 

In the calculation of the metric, where $K5 = 0$ (the default), use the following formula:
\begin{equation}
	W_1=K1*bandwidth + \frac{K2*bandwidth}{256-loading} +K3*delay
		\label{f13}
\end{equation}

If $K5$ is not equils 0, then additionally such an operation is performed:
\begin{equation}
			W=W_1*\frac{K5}{relability + K4}
			\label{f14}
\end{equation}

The final metric of Eqn.~(\ref{f14}) contains all of the same set of network variables, which is used to calculate the universal metric (\ref{f9}), and the selection of weights allows to properly comparing the metric of EIGRP from Eqns.~(\ref{f12}-\ref{f14}) in selected ranges of network variables.

\section{Summary and future research plan}
\label{s4}

The main idea of this paper is to use the mathematical apparatus of theoretical physics to describe the processes in computer networks. Information theory based Claude Shannon~\cite{sha1,sha2}, also relies on the introduction to the physics concept of entropy. The problem of determining a route similar to finding the motion trajectory, in problems of this type the variation principle is always used. Little's law from queuing theory gave a hint about the variable type for which is necessary to search for the minimum. Traffic, which is currently served by the compared network route, was selected as such a variable. The obtained formula has been upgraded with the influence of the effect of packet loss on the network hop.

Vast experience in organizing dynamic routing has been gained in operating a global network with an extremely complicated topology, which was secured in the most popular algorithms such as {\it RIP, OSPF, EIGRP}. In this paper we attempt to derive the metric functions of these algorithms, based on the found universal metric. All variables that are contained in the universal metric of Eqn.~(\ref{f9}) were used in constructing these algorithms. The experiments helped to establish the basic types of dependencies, and the metric constructed by adding terms with weighting factors to previously found terms.

The question of practical implementation of the metric and the construction of new routing algorithm are not considered in this work because it requires a separate pilot studies in order to calculate the universal metric with the lowest cost, without loading the network measurements, and not allowing the route change frequently.

We should also consider further theoretical studies, since the principle of extremes is inextricably linked with the variation principle. The main thing is to understand how to produce a variation of the variables and which effect will be to describe the resulting equation. It is likely that it will describe the features of the single router with multiple channels, in contrast to the universal metric function, which describes the extended route.


\begin{thebibliography}{99}

\bibitem{1}
B. Adams, E. Cheng, T. Fox, Interdomain Multicast Solutions Guide, Cisco Press, 2002

\bibitem{cal}
P. Calyam, M. Sridharan, W. Mandrawa, H. Schopis, Performance Measurement and Analysis of H.323 Traffic, Passive and Active Measurement Workshop (PAM), 2004

\bibitem{10}
T. Clausen, P. Jacquet, C. Adjih, A. Laouit, P. Minet, P. Muhlethaler, A. Qayyum, L. Viennot, Optimized link state routing protocol (OLSR), hal.inria.fr, 2003
 
\bibitem{3}
C. Dovrolis, P. Ramanathan, and D. Moore, Packet-Dispersion Techniques and a Capacity-Estimation Methodology, IEEE/ACM Transactions on Networking, v.12, n. 6, December 2004, p. 963-977

\bibitem{11}
P. Jacquet,P. Muhlethaler, T. Clausen, A. Laouiti, A. Qayyum, L. Viennot, Optimized link state routing protocol for ad hoc networks, IEEE INMIC, 2001

\bibitem{5}
M. Jain, C. Dovrolis, End-to-end Estimation of the Available Bandwidth Variation Range, In: SIGMETRICS'05, Ban, Alberta, Canada (2005).

\bibitem{8}
D. B. Johnson, A note on Dijkstra's shortest path algorithm, Journal of the ACM (JACM), dl.acm.org, 1973

\bibitem{knuth}
D. E. Knuth, A generalization of Dijkstra's algorithm, Information Processing Letters, Elsevier, 1977, 6(1).

\bibitem{2}
L.D. Landau and E.M. Lifshits, Theoretical Physics, Volumes 1-9, Moscow, Nauka, 1989

\bibitem{9}
M. K. Marina, S. R. Das, On-demand multipath distance vector routing in ad hoc networks, Network Protocols Ninth International Conference, ICNP, 2001

\bibitem{4}
R.S. Prasad, C. Dovrolis, and B. A. Mah, The effect of layer- 2 store-and-forward devices on per-hop capacity estimation, in Proc. IEEE INFOCOM, Mar. 2003, pp. 2090-2100

\bibitem{sag}
E.S. Sagatov, A.M. Sukhov, Duplication of Key Frames of Video Streams in Wireless Networks, IFIP Wireless Days 2011
 
\bibitem{sha1}
C. E. Shannon, Prediction and Entropy of Printed English. Bell Sys. Tech. J (3), 1950
 
\bibitem{sha2}
 C. E. Shannon, Communication in the presence of noise, Proc. Inst. Radio Engrs. N.Y., 1949 v. 37, pp. 10�21
 
\bibitem{7}
A. Sukhov, N. Kuznetsova, A. Pervitsky, and A.A. Galtsev. Generating Function For Network Delay. 2010; arXiv:1003.0190v1.

\bibitem{6}
D. Verma, H. Zhang, and D. Ferrari. Delay jitter control for real-time communication in a packet switching network, In Proccedings of TriCom '91, pp 35-43, 1991.
 
\bibitem{wpd}
D. Waitzman, C. Partridge and S. Deering, Distance vector multicast routing protocol (DVMRP), IETF RFC 1075, 1988

\end{thebibliography}
\end{document}